\documentstyle[twoside,fleqn,epsfig,espcrc2]{article}

\newcommand{\AmS}{{\protect\the\textfont2
  A\kern-.1667em\lower.5ex\hbox{M}\kern-.125emS}}

\title{A Wilson-Yukawa model with a chiral spectrum in 2D.}

\author{P. Hern\'andez\address{Theory Division, CERN,\\CH-1211 Geneva 23, Switzerland.}%
        and 
        Ph. Boucaud\address{LPTHE, Universit\'e de Paris XI\\
        91405 Orsay Cedex, France.}}
       
\begin{document}

\begin{abstract}
We summarize our recent study \cite{us3} of the fermion spectrum in a fermion-scalar 2D model with a 
chiral $U(1)_L \times U(1)_R$ global symmetry. This model is obtained 
from a two-cutoff formulation of a 2D $U(1)$ chiral gauge theory, in the limit of zero gauge coupling. 
The massless fermion spectrum found deep in the vortex 
phase is undoubled and chiral.
\end{abstract}
\maketitle

\section{Introduction}

The purpose of this study is to perform a non-trivial test of the lattice 
formulation of chiral gauge theories proposed in \cite{us1}. 
The main idea in this proposal is to reduce the unavoidable 
breaking of chiral (gauge) symmetry, by means of a regulator which separates
the scale of the fermion cutoff (where chirally breaking interactions 
become relevant) from the boson cutoff \cite{fs,us1}. In practice, 
this is achieved \cite{us1} by coupling lattice fermions on 
a lattice of spacing $f$ to gauge link variables, 
that are constructed by an appropriate 
smooth and gauge-invariant interpolation of gauge configurations 
generated on a coarser lattice of spacing $b$. As long as the 
interpolation is smooth, the Fourier modes of the interpolated field
are effectively cut off at the scale $1/b \ll 1/f$. Doublers are given a mass
by introducing a free Laplacian Wilson term.
In the absence of gauge anomalies, 
the breaking of gauge invariance is suppressed by powers of the ratio of 
the cutoff scales to all orders in the gauge coupling \cite{us1}. However, 
at any finite lattice spacing,
there is a residual breaking of the gauge symmetry and it is 
non-trivial to ensure that the continuum limit of such a theory 
has the right degrees of freedom. 
In order to address this question, it is convenient to consider the 
equivalent ``higgs'' picture in which gauge invariance is exact but there 
are extra charged scalars, which correspond to the pure gauge degrees of 
freedom \cite{fnn}.
An old and simple argument by the authors of \cite{fnn} suggests that a mild 
breaking of gauge invariance (i.e. gauge breaking interactions are characterized by some small coupling $\epsilon$) in a gauge theory
 is irrelevant at long distances. Or in other words, the theory is 
in the same universality class as the pure gauge theory. 
The reason is that, for arbitrarily small 
$\epsilon$, the correlation length of the scalar fields ($\xi$) can be
 arbitrarily small. If it is much smaller than the correlation length of the gauge-invariant degrees of freedom, the scalars can be integrated out, resulting in a local 
(within distances of $O(\xi)$) and gauge-invariant theory. In the case of 
chiral gauge theories, the gauge breaking interactions come from the 
fermionic action. The ``higgs'' picture of the two-cutoff formulation of 
\cite{us1} is, 
\begin{eqnarray}
Z = \int_{{\cal L}_b} {\cal D}\Omega {\cal D}U \; e^{-S_{g}[U]} \;e^{\Gamma[u^\omega[U]]} \; e^{-\bar{\eta} G[u^\omega[U]] \eta},
\label{tco}
\end{eqnarray}
where the $u_{\mu}[U]$ and $\omega={\cal F}[\Omega, U]$ fields are the 
interpolations to the $f$-lattice
of the $b$-lattice fields $U_{\mu}$ and $\Omega$ respectively. 
It is easy to show, using the properties of the interpolation \cite{us3}, that this action is invariant under $b$-lattice gauge transformations
$\Phi$, $U_{\mu} \rightarrow U^\Phi_{\mu}$, $u_{\mu} \rightarrow u^\phi_{\mu}$ and
$\omega \rightarrow \omega \phi^\dagger$, where $\phi \equiv {\cal F}[\Phi,U]$. In order to ensure that the continuum limit of (\ref{tco}) is a chiral 
gauge theory, it is necessary 
that the scalars decouple in this limit. For this, 
it is sufficient that the interactions 
of the scalars induced by $\Gamma[u^\omega]$ are small in the sense of \cite{fnn}. That this is the case for an anomaly free theory in the two-cutoff 
formulation
was shown in \cite{us1}, with $\epsilon = O(f/b)^2$. Thus, for small enough
$f/b$, the $\Omega$ fields should decouple.
Besides, one should also make sure that the fermion spectrum is the correct
one, after the $\Omega$ fields are integrated out. In \cite{testa}, 
it has been shown that, in 
the one-cutoff version of (\ref{tco}) (i.e., $f=b$ and
$\omega = \Omega$ and $u_{\mu}= U_{\mu}$), the $\Omega$ integration 
cancels the effects of the Wilson term and there is doubling. 
Although, the argument of \cite{testa} is not applicable for $f/b \neq 1$, 
it is important to study the fermion spectrum of (\ref{tco}), since 
the scalar-fermion dynamics is non-peturbative. If there was a 
doubling problem, this could be detected through the effective action,
\begin{eqnarray}
e^{\Gamma^{eff}[U]} \equiv \int_{{\cal L}_b} {\cal D}\Omega \;\;e^{\Gamma[u^\omega[U]]},
\label{effact}
\end{eqnarray}
or/and the fermion propagator, 
\begin{eqnarray}
G^{eff}[U] = \int_{{\cal L}_b} {\cal D}\Omega \;\;e^{\Gamma[u^\omega[U]]} \;\;G[u^\omega[U]].
\label{effprop}
\end{eqnarray}
In this work \cite{us3}, we are concerned with the fermion propagator. 
We have computed (\ref{effprop}) in the simplest non-trivial model,
which is a U(1) chiral theory in 2D, 
in the quenched approximation, and in the global limit (i.e. the gauge coupling is set to zero).
In this limit, we should find free, massless and undoubled
fermions with chiral quantum numbers under the residual global symmetry, if
(\ref{tco}) is to describe a chiral gauge theory. 

\section{Fermion Spectrum}

We refer the reader to \cite{us3} for details
about the two-cutoff formulation of this model. The propagator is given by $G[u^\omega[U]] \equiv (\hat D)^{-1}$, with
\begin{eqnarray}
{\bar \Psi} {\hat D} \Psi  \equiv \bar{\Psi}  
[\frac{\gamma_{\mu}}{2} (( D^+_{\mu} + D_{\mu}^-) L + (\partial^+_{\mu}+\partial^-_{\mu}) R) \nonumber\\
 + y (\omega^\dagger R + \omega L ) - \frac{r}{2} (\omega^\dagger \partial_{\mu}^+ \partial^-_{\mu} R  
+ \partial_{\mu}^+ \partial^-_{\mu} \omega L )]\Psi,
\label{u1}
\end{eqnarray}
where the covariant and normal derivatives are given by $D_{\mu}^+ \Psi(x) = u_{\mu}(x) \Psi(x+\hat{\mu}) - \Psi(x)$, $D_{\mu}^- \Psi(x) = \Psi(x) - u^\dagger_{\mu}(x-\hat{\mu}) \Psi(x-\hat{\mu})$, $\partial_\mu^+ = D_{\mu}^+|_{u=1}$ and $\partial_\mu^- = D_{\mu}^-|_{u=1}$. $L, R$ are the left and right chirality
projectors. In the global limit, $u=1$ in the
previous formulae and the $\omega$ fields depend only on $\Omega$ (the 
detailed formulae can be found in the appendix of \cite{us3}). In this limit, 
the action (\ref{u1}) has a global $U(1)_R\times U(1)_L$. The correct 
spectrum should consist of two undoubled 
massless fermions with charges $(q_R, q_L)= (0,1), (1,0)$.

The fermionic action (\ref{u1}) 
is a two-cutoff version of the action studied in the 
context of Wilson-Yukawa (WY) models in \cite{smit}\footnote{The vortex phase of the WY model of \cite{smit} is obtained by setting $f/b=1$, $\omega = \Omega$ in (\ref{u1}) and adding a kinetic term for $\Omega$ with 
strength $\kappa\leq \kappa_c$.}. In those studies, two phases were found on the plane ($y$, $r$), in the vortex phase of the scalar dynamics.
 A strong Yukawa phase for $d r + y \geq 1$ ($d$ is the dimension of space-time), where, due to the strong fermion-scalar coupling, the so-called neutral composite Dirac
field $\Psi^{(n)} \equiv \omega \Psi_L + \Psi_R$ forms and gets a mass. The resummation of the first order corrections in the strong 
coupling expansion gives $m^{(n)} \sim z^{-1}$, where $z^2 \equiv < Re[ \omega(x)  \omega^\dagger(x+\hat \mu) ] >$  is a chirally-invariant condensate which does not vanish in the vortex phase. The charged fermion 
$\Psi^{(c)} \equiv \Psi_L + \omega^\dagger \Psi_R$ is a two particle state, $\omega \Psi^{(n)}$, and 
no other fundamental fermion with $U(1)_L$ charge was found in the spectrum. For
perturbative Yukawa couplings ($d r + y \leq 1$), on the other hand, all the fermion masses were found to be $\sim v \equiv < \omega >$, which vanish
in the vortex phase, so doublers were present. Neither the strong nor 
the weak phase could then give rise to a chiral gauge theory in these
models. On the other hand, in \cite{us3}, we have found that for $f/b << 1$ a new truly chiral phase exists for large $r$ and small $y$. 

For each random $\Omega$ configuration, the $\omega$ fields are obtained
from the interpolation of \cite{us3}, and the operator $\hat D$ is 
inverted using Conjugate Gradient (CG). By choosing the appropiate 
sources, the neutral and charged fermion propagators, $S^{(n,c)}$, are 
also computed. An average is obtained from a sample
of 200-500 configurations, depending on the value of $y$.
The expectation value of $\omega$, $v$, vanishes, as expected, since we are in the vortex phase. However, the chirally symmetric condensate $z^2$ is non-zero.
Since we are interested in decoupling the doublers, we 
have set $r=1$ through out and varied $y$. The spectrum obtained differs 
at large and small $y$. At large $y$, our results
are similar to those found in the strong phase of 
WY models. In particular, the neutral propagator is well described
by the strong Yukawa coupling expansion \cite{us3}.
In contrast with WY models, however, 
as $y$ is decreased, the components $S^{(n)}_{RL}$ and $S^{(c)}_{LR}$ 
become lighter and, at some finite $y_c$, develop poles at zero momentum. 
\begin{figure}
\begin{center}
\mbox{\epsfig{file=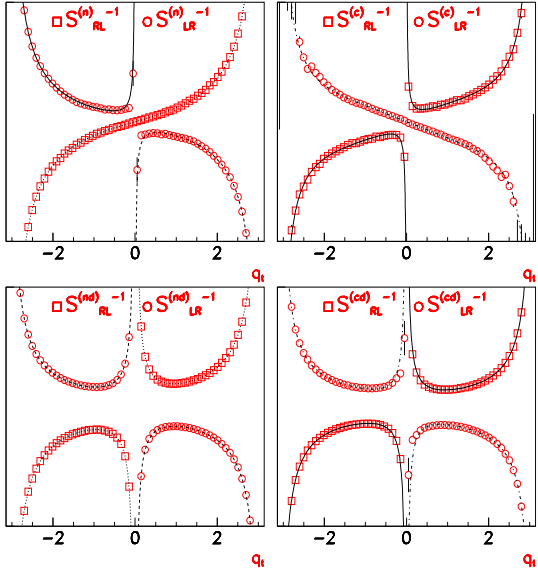,width=2.8in,height=3.in}}
\end{center}
\vspace{-15pt}
\caption[]{$(S^{(n)}_{RL,LR}(q_t))^{-1}$, $(S^{(c)}_{RL,LR}(q_t))^{-1}$, $(S^{(nd)}_{RL,LR}(q_t))^{-1}$, $(S^{(cd)}_{RL,LR}(q_t))^{-1}$ for $\frac{L}{b}_{\frac{b}{f}}=8_8$ and $y=0.05$.}
\label{fig1}
\vspace{-25pt}
\end{figure}
Fig. 1 shows the components $(S^{(n,c)}_{RL,LR})^{-1}$  for $q_x=0$ 
as a function of the 
temporal momentum and $(S^{(nd,cd)}_{RL,LR})^{-1}$ for $q_x=\pi$, corresponding to the spatial doublers. The components describing the propagation of $\Psi^{(n)}_R= \Psi_R$ and $\Psi^{(c)}_L = \Psi_L$ have simple
poles at $q_t=0, q_x=0$ and no other poles in the Brillouin zone. These are 
the two expected massless fermions with $(q_R, q_L) = (1,0),(0,1)$. The other components of the neutral and charged 
propagators, $S^{(n)}_{LR}$, $S^{(c)}_{RL}$, have no poles in the Brillouin zone. At small momemtum they behave 
as the two particle states $\omega \Psi_L$ and $\omega^\dagger \Psi_R$ respectively. At large momenta, on the other hand, the neutral field behaves  
as a massive Dirac field, in agreement with the strong coupling expansion \cite{us3}. This can be seen in Fig.1 from the behaviour of the spatial doublers 
or from the behaviour of $S^{(n)}$ near $q_t = \pi$. In summary, for 
$y\leq y_c$ and
$r=1$, the massless fermion spectrum is undoubled and chiral as expected.
Naive power counting qualitatively explains this new chiral phase. The couplings of the 
light modes and the scalars are suppressed, either by $y$ or by $f/b\; d r$, so for small enough $f/b$ and $y$, at fixed $r$, 
the behaviour of propagators near zero momemtum
should be as in the weak phase. On the other hand, the
 coupling of the doublers to the scalars is mainly
 controlled by $d r$, so the propagators near the doubler
momenta behave as in the strong phase, for $d r \geq 1$. 

There is some numerical evidence for the naive guess $y_c \sim f/b$,
 coming from the position of the maximum of the number of 
CG iterations in the fermion matrix inversion \cite{us3}. We have not explored the boundaries of the chiral phase as 
a function of $r$. But clearly at small $d r \leq 1$ we should get back
 to the perturbative regime, 
while for large $d r \sim b/f$, the chiral phase should be lost because the coupling
of the light modes to the scalars becomes strong. This also suggests
what happens when $f/b \rightarrow 1$. Since the chiral phase is expected to 
exist in some band 
$0 < y < y_c$ and $d^{-1} < r < d^{-1} b/f$, it gets squeezed to zero as 
$f/b \rightarrow 1$. Unfortunately, this also implies that there is no simple
 analytical expansion to study this phase, since $r$ cannot be treated neither as small nor as large.

\end{document}